\documentclass[aps,pra,twocolumn,final]{revtex4-1}

\usepackage{epstopdf}
\usepackage{graphicx}

\graphicspath{{figs/}}
\usepackage{amssymb} 
\usepackage{mathtools}
\usepackage{mathrsfs}
\usepackage[usenames]{color} 
\usepackage{textcase}
\usepackage{bm}

\usepackage{subcaption}
\usepackage{setspace}
\usepackage{natbib}
\usepackage{bbold}
\newcommand{\ignore}[1]{}
\usepackage{cleveref}
\begin{document}
\title{Strong field ionization and gauge dependence of nonlocal potentials}

\author{T.C. Rensink}\affiliation{University of Maryland, College Park}
\author{T.M. Antonsen Jr.}\affiliation{University of Maryland, College Park}

\date{9/6/2016}

\begin{abstract}
Nonlocal potential models have been used in place of the Coulomb potential in the Schrodinger equation as an efficient means of exploring high field laser-atom interaction in previous works.  Although these models have found use in modeling phenomena including photo-ionization and ejected electron momentum spectra, they are known to break electromagnetic gauge invariance.  This paper examines if there is a preferred gauge for the linear field response and photoionization characteristics of nonlocal atomic binding potentials in the length and velocity gauges.  It is found that the length gauge is preferable for a wide range of parameters.
\end{abstract}

\maketitle	
\section{INTRODUCTION}
Study of strong field ultra-short pulse laser gas interactions, including as THz frequency radiation generation \citep{Kim2007,Johnson2013,Chen2015}, high harmonic generation \citep{Lewenstein1994}, and the growing field of attosecond atom-field dynamics \citep{Dahlstrom2013}, relies on numerical modeling of laser-gas interaction.  This is often done in two different regimes:  A ``macroscopic'' simulation of laser-pulse evolution over distances of millimeters or centimeters, where the gas is treated as a medium that includes the linear field response, nonlinear field response, including the possibly rotational field response for a diatomic gas, field ionization, and free electron response \citep{Couairon2011, Kolesik2016}.  In the second, ``microscopic'', regime the interaction of the field with a single atom or molecule is examined in the quantum mechanical picture.  This in principle requires solution of the time dependent Schrodinger equation (TDSE) using approximate analytical methods \citep{Popruzhenko2008}, finite-difference time domain (FDTD) numerical solutions \citep{Gordon2012}, or by Floquet expansion schemes \citep{Potvliege1998}.  Although attempts have been made to couple Maxwell's Equations with a ``microscopic'' Schrodinger model \citep{Lorin2012}, these simulations are computationally expensive and remain largely beyond reach at the time of this writing.

Nonlocal binding potentials are a promising tool for efficient solution of the Schrodinger equation, capable of reproducing many the basic quantum mechanical atomic properties efficiently.  Despite these successes, it is known that nonlocal models are gauge dependent, while classical electromagnetic theory and the Schrodinger formulation of quantum mechanics are well known to be gauge independent \citep{Han2010}.  Breaking this symmetry raises the natural question of how to handle the gauge dependence of these potentials.  

This paper examines the gauge dependence of a nonlocal gaussian potential representing the Coulomb potential in a hydrogen-like atom, in the presence of a time varying, spatially uniform electric field.  Specifically, we consider the linear polarizability and photoionization rates predicted by the nonlocal models in the length and velocity gauges.  The paper is organized as follows: Section \ref{sec:GAUGE} briefly reviews the statement of gauge invariance of the Schrodinger equation for local potentials, section \ref{sec:LG_VG} introduces the nonlocal potential formulation in the length and velocity gauges, section \ref{sec:time_indep} reviews some of the basic time-independent properties of the gaussian nonlocal model, and  section \ref{sec:time_dep} examines the static and dynamic atomic polarizability and photoionization characteristics for each gauge.  Concluding remarks follow. 

\section{Gauge Invariance of Local Potentials}\label{sec:GAUGE}

We briefly examine the gauge invariance of local potential formulations of the time dependent Schrodinger equation.  Specifically, we consider the TDSE for the wavefunction of a single electron in the presence of an atomic potential $V(\mathbf x)$ and a classical electromagnetic field in the dipole approximation with no back-reaction.  The time-dependent electric field is represented in the Schrodinger equation via the electromagnetic potential terms, defined through the relation $\mathbf F(t) = -\partial_t\mathbf A(t) - \nabla \Phi(\mathbf x, t)$, noting that, for simplicity we require $\mathbf A(t)$ depend only on time and that $\Phi(\mathbf x,t)$ be linear in $\mathbf x$.   The magnetic field is ignored.   Atomic units (a.u.) $\hbar = m_e = 1$, $q_e = -1$ are used throughout except where noted.  The general form of the Schrodinger equation is then:
\begin{equation}\label{fullTDSE}
i\partial_t\psi(\mathbf x, t)=\left[\frac{1}{2} \left(-i\nabla+\mathbf A(t)\right)^2-\Phi(\mathbf x,t)-V(\mathbf x)\right]\psi(\mathbf x ,t).
\end{equation}

The choice of $\mathbf A$ and $\Phi$ is not unique; one may define a new set of potentials $\mathbf A', \mathbf \Phi'$ with the addition of a gauge term
\begin{align}
&\mathbf A'(t) \equiv \mathbf A(t) + \nabla \chi(\mathbf x,t)\\
&\Phi'(\mathbf x,t) = \Phi(\mathbf x,t) - \partial_t \chi(\mathbf x,t)
\end{align}
that produce the same field $\mathbf F(t)$, noting that the gauge term takes the form $\chi(\mathbf x,t)= \mathbf x \cdot \Delta \mathbf A(t)$ for this system.

On defining a new wavefunction that is modified by a local phase factor,
\begin{align}\label{eq:psi_trans}
\psi'(\mathbf x,t) &= \exp\left[-i\chi(\mathbf x,t)\right]\psi(\mathbf x,t),
\end{align}

we express the original Schrodinger equation in terms of the primed variables, and operate on the gauge term, i.e. $i\partial_t \psi = \exp(i \chi) (i \partial _t - \partial_t \chi) \psi'$, and $(-i\nabla + \mathbf A(t))\exp(i \chi) \psi' = \exp(i \chi) (-i \nabla + \mathbf A(t) + \nabla \chi)\psi'$,

leading to a Schrodinger equation of equivalent form in the transformed variables
\begin{equation}\label{eq:transformed_TDSE}
i\partial_t\psi'(\mathbf x, t)=\left[\frac{1}{2} \left(-i\nabla+\mathbf A'(t)\right)^2-\Phi'(\mathbf x,t)-V(\mathbf x)\right]\psi'(\mathbf x ,t).
\end{equation}

Both the original and gauge-transformed Schrodinger equations reproduce the same set of observables and are therefore said to be gauge invariant.

\section{Gauge Dependence of Nonlocal Potentials}\label{sec:LG_VG}
If we allow the potential to take the form of an operator acting on the the wavefunction $V(\mathbf x) \psi(\mathbf x,t) \rightarrow \hat V \psi(\mathbf x,t)$, we may define a nonlocal potential \citep{TetchouNganso2007, TetchouNganso2011, TetchouNganso2013, Rensink2014} as:
\begin{align}
\hat V \psi(\mathbf x,t) &\equiv V_0 u(\mathbf x) S(t) \label{NL_def}\\
S(t) &\equiv \int d^3 \mathbf x' u^*(\mathbf x') \psi(\mathbf x',t) \label{S_def}\\
u(\mathbf x) &=  \sigma^{-3}\exp\big(-\mathbf x^2/(2\sigma^2)\big) \label{u_def}
\end{align}

where we have chosen to use a gaussian shape function for $u(\mathbf x)$.  Specifically, the nonlocal potential term is comprised of the function $u(\mathbf x)$ scaled by the projection of the wavefunction onto $u^*(\mathbf x)$.  Projecting onto the complex conjugate ensures the non-local potential remains self-adjoint.  Loosely speaking, the positive real valued constant $V_0$ controls the ``strength'' of the potential ($V_0 > 0$ is attractive) and $\sigma$, with dimension of length, controls the width of the potential.  On performing the same gauge transformation as done in the previous section (and dividing through by an overall phase factor $\exp(i\chi)$) the nonlocal potential term appears in the gauge-transformed Schrodinger equation as:
\begin{equation}\label{eq:nonlocaltransformed}
\hat V \psi(\mathbf x,t) \rightarrow \exp(-i \chi)u(\mathbf x) \int d^3 \mathbf x' u^*(\mathbf x') \exp(i \chi) \psi'(\mathbf x',t)
\end{equation}
and it can be seen that the potential term is modified by the phase factor $\chi$.  

A form of gauge invariance can be introduced if we treat $u(\mathbf x)$ as a field that undergoes the same transformation as $\psi(\mathbf x,t)$, namely $u'(\mathbf x,t) \equiv \exp(-i \chi)u(\mathbf x)$; the transformed Schrodinger equation is of the same form as the original and will yield the same observables.  However, this implies that $u(\mathbf x)$ depends on the gauge, and $u'(\mathbf x,t)$, which represents the atomic potential, now depends on the introduced field.  This is unphysical, so the question naturally arises: is there a natural gauge for introducing a nonlocal potential?  We examine two obvious choices, setting either $\mathbf A=0$ or $\Phi=0$ in Eq.\eqref{fullTDSE}, defining the electric field through a single potential term.

The analysis in the remainder of this paper will be done in the k-space (momentum) representation for convenience via the Fourier transform definitions,
\begin{subequations}
\begin{align}
\phi (\mathbf k) &= \frac{1}{(2\pi)^{3/2}}\int \!d^3  \mathbf x' \ e^{-i \mathbf k \cdot \mathbf x' } \psi(\mathbf x')\\
\psi (\mathbf x) &= \frac{1}{(2\pi)^{3/2}}\int \!d^3  \mathbf k' \ e^{i \mathbf k' \cdot \mathbf x } \mathscr \phi(\mathbf k')
\end{align}
\end{subequations}
so that the (canonical) momentum is given by $-i \nabla \rightarrow \mathbf k$.
We examine the Schrodinger equation in the so-called length gauge, where $\mathbf A(t)=0$ in Eq.\eqref{fullTDSE}, and the velocity gauge, where $\Phi(\mathbf x,t) =0$.   The momentum-space equations in these two cases are:

\begin{subequations}
\begin{align}
&\left[ i \partial_t -\frac{1}{2} \mathbf k^2 + i \partial_t \mathbf A(t) \cdot \nabla_{\mathbf k} \right] \phi_L(\mathbf k,t) = - \hat V \phi_L(\mathbf k,t) \label{eq:lgTDSE}, \\
&\text{and}\notag \\ 
&\left[ i \partial_t -\frac{1}{2}\left(\mathbf k + \mathbf A(t)\right) ^2 \right] \phi_V(\mathbf k,t) = - \hat V \phi_V(\mathbf k,t), \label{eq:vgTDSE}
\end{align}
\end{subequations}

where the subscripts designate length and velocity gauge wavefunctions respectively.  The nonlocal potential operator is identical in both equations, specifically 
\begin{align}
&\hat V \phi(\mathbf k,t) \equiv V_0u(\mathbf k) \int d^3 \mathbf k' u^*(\mathbf k') \phi(\mathbf k',t)\\
&u(\mathbf k) \equiv \exp(- \sigma^2 \mathbf k ^2 /2).
\end{align}
We note that the electric potential is written in terms of a single variable $\mathbf A(t)$ in both equations, where the electric field is defined as $\mathbf E(t) = -\partial_t\mathbf A(t)$.  

Although we have represented the electric field using a common potential, Eqs.\eqref{eq:lgTDSE}, \eqref{eq:vgTDSE} are not equivalent.  We substitute the explicit expressions for the nonlocal potential in Eqs.\eqref{eq:lgTDSE} and \eqref{eq:vgTDSE}, introduce integrating factors, and obtain:

\begin{subequations}
\begin{align}
 \phi_{L}(\mathbf k,t) &= i V_0\int \limits _{-\infty}^t dt' \exp\biggl[-\frac{i}{2} \int _{t'}^{t} dt'' \left(\mathbf k - \mathbf A(t) + \mathbf A(t'') \right)^2 \biggl]\ldots \notag \\
 &u(\mathbf k - \mathbf A(t)+ \mathbf A(t'))S_L(t') \label{eq:phi_L}\\ 
\notag\\
\phi_{V}(\mathbf k,t) &= i V_0\int \limits _{-\infty}^t dt' \exp\left[-\frac{i}{2} \int _{t'}^{t} dt'' \left(\mathbf k + \mathbf A(t'') \right)^2 \right] \ldots \notag \\
&u(\mathbf k)S_V(t') \label{eq:phi_V}
\end{align}
\end{subequations}

where 
\begin{equation}
S_{V,L} = \int d^3 \mathbf k u^*(k) \phi_{V,L},
\end{equation}
which follows from Eq.\eqref{S_def}.

\section{FIELD FREE SYSTEM}\label{sec:time_indep}
If $\mathbf A(t) = 0$, the length and velocity gauge systems are equivalent.  The time independent system is found to have a single bound state which can be represented explicitly in momentum space:

\begin{align}\label{eq:gs_profile}
\phi_0(\mathbf k) &= \frac{2V_0 S_0 u(\mathbf k) }{|\mathbf k|^2 + k_0^2}\\ 
S_0 &\equiv \int d^3 \mathbf k' u^*(\mathbf k') \phi_0(\mathbf k')
\end{align}
where $k_0 \equiv \sqrt{2 E_0}$ is real and positive defined for a state with total energy $-|E_0|$.  Multiplying both sides of Eq.\eqref{eq:gs_profile} by $u^*(\mathbf k)$, integrating over all momenta, and dividing both sides by $S_0$,  we obtain the consistency relation
\begin{equation}\label{eq:disp1}
1 =  2V_0 \int d^3 \mathbf k' \frac{ |u(\mathbf k')|^2 }{|\mathbf k'|^2 + k_0^2}.\\ 
\end{equation}
Equation \eqref{eq:disp1} relates $E_0$, $\sigma$, and $V_0$, which can be integrated to give
\begin{align}\label{eq:dispersion}
1 =  \frac{4 \pi ^{3/2}V_0}{\sigma} \left[ 1 - \sqrt{\pi}\sigma k_0 \exp\left( \sigma^2  k_0^2 \right) \text{erfc}\left( \sigma  k_0  \right)\right].
\end{align}
Here, erfc is the complimentary error function.  For a system with a single nonlocal binding potential term, equation \eqref{eq:disp1} implies that only a single bound state is supported by the nonlocal potential (in contrast to a gaussian local potential \citep{Lai83}); for a chosen value of $V_0$ and $\sigma$, only a single value of $E_0 = k_0^2/2$ will satisfy the consistency relation Eq. \eqref{eq:disp1}.  

\begin{figure}
\includegraphics[width=0.5\textwidth]{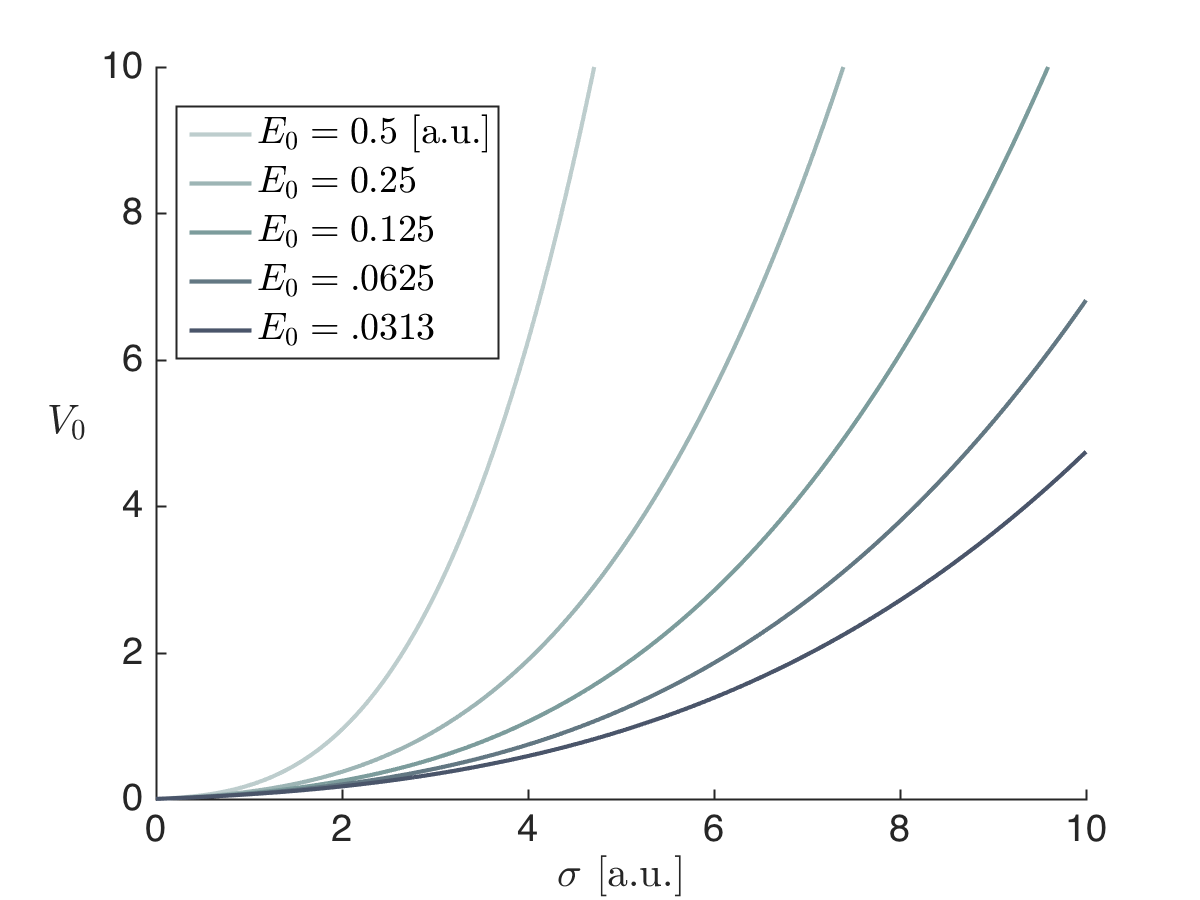}
\caption[justification=justified]{(Color online) Curves relating $V_0$ and $\sigma$ for constant values of $E_0$ that satisfy Eq.\eqref{eq:dispersion}.  Shown here for the first five hydrogen states, the gaussian nonlocal potential supports a (single) bound state of arbitrary energy.}
\label{fig:EV}
\end{figure}

\begin{figure}
\centering
{\includegraphics[width=8.5cm]{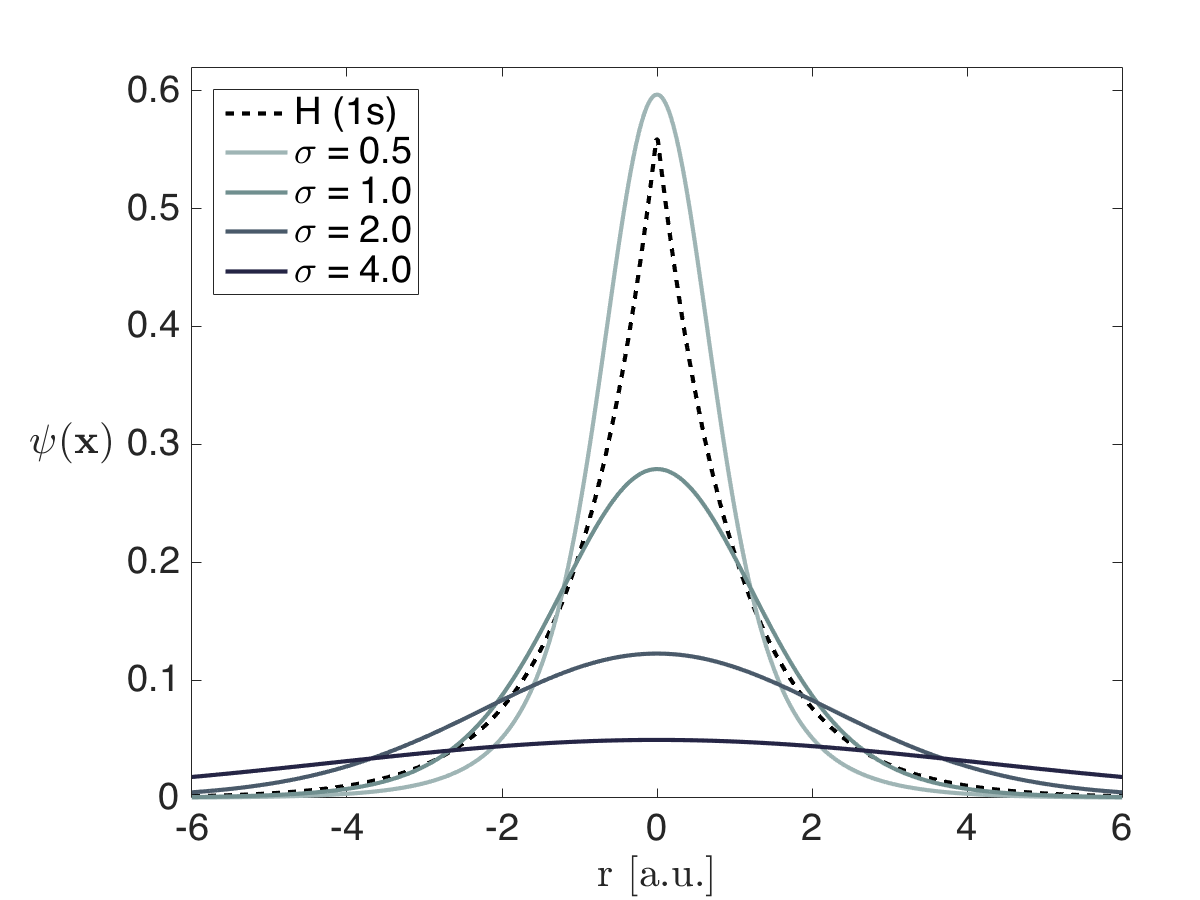}}
\caption{(Color online) The normalized configuration space wavefunction $\psi(\mathbf |\mathbf r|)$ is given by the Fourier transform of Eq.\eqref{eq:gs_profile} (shown here for $E_0 =.5$).   The variable $\sigma$ is used as a fitting parameter.}
\label{fig:gs_profile}
\end{figure}

Figure \ref{fig:EV} shows the values of $V_0$ vs $\sigma$ for the energies corresponding to the first five states of the hydrogen eigen-spectrum, $E_n = .5/n^2$.  Once the bound state energy is specified, $\sigma$ is used as a fitting parameter that determines $V_0$ via Eq.\eqref{eq:dispersion}.  Figure \ref{fig:gs_profile} shows a comparison of the nonlocal wavefunction, $\psi_0(\mathbf x)$, for various values of $\sigma$ ($E_0 = .5$); the hydrogen 1s orbital is provided for comparison.  

\section{Field response in the Length and Velocity Gauges}\label{sec:time_dep}

Equations \eqref{eq:phi_L} and \eqref{eq:phi_V} show that the time-dependent wavefunction, in the presence of a time varying field, can be recovered if the (gauge dependent) overlap functions $S_L(t)$, $S_V(t)$ are known.  These in turn depend on integrals of the wavefunctions (see Eq.\eqref{S_def}).  The advantage of the nonlocal potential model is that these integrals can be carried out analytically, resulting in Volterra (type II) integral equations for the functions $S_{L,V}(t)$.  

This method reduces a 3+1 dimensional calculation of $\psi(\mathbf x,t)$ typically needed to find values of the wavefunction and observables of interest to a series of $\sim$ 2D calculations (the number of operations required to solve the integral equation in time grows like $t^2$).  Further, since the wavefunction has been integrated analytically, the approach is not limited by spatial (or momentum) resolution or extent, which can present difficulties for finite difference solvers.  Loosely speaking, the spatial/momentum dependence has been ``integrated out'' while encoding the wavefunction evolution through the time evolution of the complex variable $S(t)$.  

The integral equation for $S_{L,V}(t)$ is found by multiplying Eqs.\eqref{eq:phi_L}, \eqref{eq:phi_V} by $u(\mathbf k)$ and integrating over all momenta.  The resulting equation can be written in terms of a kernel function that depends on known quantities:
\begin{align}
\label{eq:S_integral}
S_{L,V}(t) = \int\limits_{-\infty}^t dt' K_{L,V}(t,t') S_{L,V}(t').
\end{align}
The kernel $K_{L,V}$ is different in the length and velocity gauges:
\begin{subequations}
\begin{align} \label{eq:K_equations}
&K_L =iV_0 \left[\frac{2 \pi}{\alpha(t,t')}\right]^{3/2} \ldots \\
&\exp\left[- \sigma^2(\mathbf A^2 + \mathbf A'^2) + \frac{1}{2 \alpha(t,t')}\big(i \Delta \mathbf x + \sigma^2(\mathbf A + \mathbf A')^2\big) \right], \notag\\
\notag \\
&K_V = iV_0 \left[\frac{2 \pi}{\alpha(t,t')}\right]^{3/2} \exp\left[\frac{i \Delta \mathbf x}{2 \alpha(t,t')} \right],
\end{align}
\end{subequations}
where
\begin{align}
&\alpha(t,t') \equiv 2\sigma^2 + i(t-t') \label{eq:alpha_t},
\end{align}
and
\begin{align}
&\Delta \mathbf x \equiv \int \limits_{t'}^t \mathbf A(t'') dt'' = \mathbf x(t) - \mathbf x(t') \label{eq:Delta}.
\end{align}
The variable $\Delta \mathbf x$ corresponds to the displacement of a classical electron in the presence of $\mathbf A$ from time $t'$ to $t$ (assuming the initial velocity $\mathbf v(t') = 0$).  In obtaining \eqref{eq:S_integral}-\eqref{eq:Delta}, we have absorbed an overall spatially independent phase factor $\exp(\int_{t'}^t dt'' \mathbf A^2(t''))$ into the definition of the wavefunction, which will not affect any results.  The velocity gauge and length gauge kernels differ due to he explicit appearance of the potential, $\mathbf A(t), \mathbf A(t')$ in the length gauge kernel; all the field-dependence in the velocity gauge expression appears through the variable $\Delta \mathbf x$ (as was true for Eqs.\eqref{eq:phi_L}, \eqref{eq:phi_V}).

\subsection{Atomic Dipole Moment}
The average momentum and time dependent atomic dipole moment are defined as
\begin{equation}\label{eq:k_ave}
\langle \mathbf k \rangle \equiv \int d^3 \mathbf k' \phi^*(\mathbf k',t) \mathbf k' \phi(\mathbf k',t)
\end{equation}
and
\begin{align}\label{eq:dipole}
\mathbf p(t) &\equiv -\int d^3 \mathbf x' \psi^*(\mathbf x',t) \mathbf x \psi(\mathbf x',t)\notag \\
&= -i\int d^3 \mathbf k' \phi^*(\mathbf k',t) \nabla \phi(\mathbf k',t)
\end{align}

In principle, the nonlinear dipole moment, including the effects of ionization, can be determined from the wavefunction given as the solution of Eqs.\eqref{eq:phi_L}, \eqref{eq:phi_V}.  However, as shown in \citep{Rensink2014}, it is computationally less intensive to solve for the dipole moment using the Ehrenfest relations.  These are written as two first-order coupled ODE's with integral expressions for $S(t)$.  In both length and velocity gauges:
\begin{align}\label{eq:k}
\partial_t \langle \mathbf k \rangle &= 2\text{Im}\left[V S^*(t) \int \limits _{-\infty}^t \mathbf M(t,t') S(t') \right]\\
\partial_t \langle \mathbf p \rangle &= -\langle \mathbf k \rangle - \mathbf A(t) + 2\text{Re}\left[V S^*(t) \int \limits _{-\infty}^t \mathbf L(t,t') S(t') \right]
\end{align}
provided we use different definitions for the kernel terms $\mathbf L, \mathbf M, \mathbf n$,
\begin{align}
\mathbf L_{L}(t,t') &\equiv \sigma^2\Big(\mathbf n_{L}(t,t') - \mathbf A(t)\Big) K_{L}(t,t')\\
\notag \\
\mathbf M_{L}(t,t') &\equiv -\Big(\mathbf n_{L}(t,t') - \mathbf A(t) \Big) K_{L} (t,t')\\
\notag \\
\mathbf n_{L}(t,t') &\equiv \frac{i \Delta \mathbf x(t,t') + \sigma^2 \Big(\mathbf A(t) + \mathbf A(t')\Big)}{\alpha(t,t')}
\end{align}
and
\begin{align}
\mathbf L_{V}(t,t') &\equiv \sigma^2 \mathbf n_{V}(t,t') K_{V}(t,t')\\
\notag \\
\mathbf M_{V}(t,t') &\equiv -\mathbf n_{V}(t,t') K_{V} (t,t')\\
\notag \\
\mathbf n_{V}(t,t') &\equiv \frac{i \Delta \mathbf x(t,t') }{\alpha(t,t')}
\end{align}
where subscript $L,V$ indicate the length and velocity gauges respectively, using previous definitions for $\Delta \mathbf x$, $\alpha$, and $\sigma$ in Eqs.\eqref{eq:alpha_t}, \eqref{eq:Delta}.  The velocity gauge expressions are again reductions of the length gauge expression where explicit appearances of the potential $\mathbf A(t)$ and $\mathbf A(t')$ are absent.  

\subsection{Linear Polarizability}
In the low field regime, the (total) dipole moment in Eq.\eqref{eq:dipole} can be characterized by the frequency dependent polarization
\begin{equation}\label{eq:alpha}
\mathbf {\hat p}(\omega) = \alpha(\omega) \mathbf {\hat F}(\omega)
\end{equation}
where $\alpha(\omega)$ is the dynamic polarizability.  Although generally a tensor, $\alpha(\omega)$ can be represented here by a scalar function because the nonlocal potential is isotropic in $\mathbf k$, $\mathbf x$ and is related to the electric susceptibility tensor  $\chi^{(1)}(\omega)$ through the Clausius-Mossotti relation (\citep{Boyd2007}).

To obtain the expression for $\alpha(\omega)$ for the nonlocal potential model, we define the following: 
\begin{align*}
\mathbf {F}(t) &=  \hat{\mathbf F} e^{-i \omega t} + c.c.,\\
\mathbf {A}(t) &= \frac{ \hat{ \mathbf F}}{i \omega} e^{-i \omega t} + c.c.,\\
\mathbf {p}(t) &= \mathbf { \hat p} e^{-i \omega t} + c.c.,\\
\phi(\mathbf k,t) &\rightarrow \left( \phi_0(\mathbf k) + \delta \phi(\mathbf k,t)\right) e^{iE_0 t},\\
\delta \phi(\mathbf k,t) &\equiv \phi_-(\mathbf k)e^{-i \omega t} + \phi_+(\mathbf k)e^{i \omega t}, \\
S_0 &\rightarrow \left( S_0 + \delta S(t) \right)e^{iE_0t},\\
\text{and} \quad & \notag \\
\delta S(t) &\equiv \int d^3 \mathbf k' u^*(\mathbf k') \delta \phi(\mathbf k',t) = S_-e^{-i\omega t} + S_+e^{i\omega t}, 
\end{align*}
where $\omega$ is the frequency of the applied field, and we require $\mathbf {F}(t)$, $\mathbf {A}(t)$, and $\mathbf p(t)$ to be real quantities.  The expressions above are inserted in a perturbative expansion of the Schrodinger equation (Eq. \eqref{eq:vgTDSE}) and solved for $\delta \phi$ (discarding all higher order terms).  The result is used in Eq.\eqref{eq:dipole} to obtain the first order, frequency dependent dipole moment.  For a linearly polarized monochromatic field $\mathbf  F(t) =  F_0 e^{-i \omega t}\mathbf {\hat z}$, one obtains the following for the velocity gauge treatment:
\begin{align*}
\phi_- &= D(-\omega)\left[V_0 u(\mathbf k) S_{-} - \phi_0 \frac{k_z}{i\omega}  \hat{ F}\right],\\
\phi_+&= D(\omega)\left[V_0 u(\mathbf k) S_{+} + \phi_0 \frac{k_z}{i\omega} \hat{ F}^* \right],\\
\text{where} \quad \notag \\
D(\pm \omega) &\equiv \left(E_0 + \mathbf k^2/2 \pm \omega \right)^{-1}.\\
\end{align*}
With some algebraic manipulation one finds expressions for $\mathbf {\hat p_{\pm}}$, e.g:
\begin{align*}
\hat{\mathbf p}_- = 
-&\int d^3 \mathbf k \ D(-\omega)\phi_0 \left(\partial_{k_z} \phi_0 \right)\frac{ k_z}{\omega}  {\hat F} \ +\\
&\int d^3 \mathbf k \ D(\omega) \phi_0 \left(\partial_{k_z} \phi_0\right) \frac{ k_z}{\omega} {\hat F}.
\end{align*}
A similar expression can be found for $\mathbf p_+$.  The polarizability is then given by the expression:

\begin{align}\label{eq:VG_pol}
\alpha_V(\omega) = \int d^3 \mathbf k \left[D(\omega)  - D(-\omega) \right] \phi_0 \left(\partial_{k_z}\phi_0\right) \frac{ k_z }{\omega},
\end{align}
and the length gauge polarizability is found by the same method to be
\begin{align}
\label{eq:LG_pol}
\alpha_L(\omega) = \int d^3 \mathbf k \left[D(\omega) + D(-\omega) \right](\partial_{k_z} \phi_0)^2.
\end{align}

\begin{figure}
\centering
{\includegraphics[width=8.5cm]{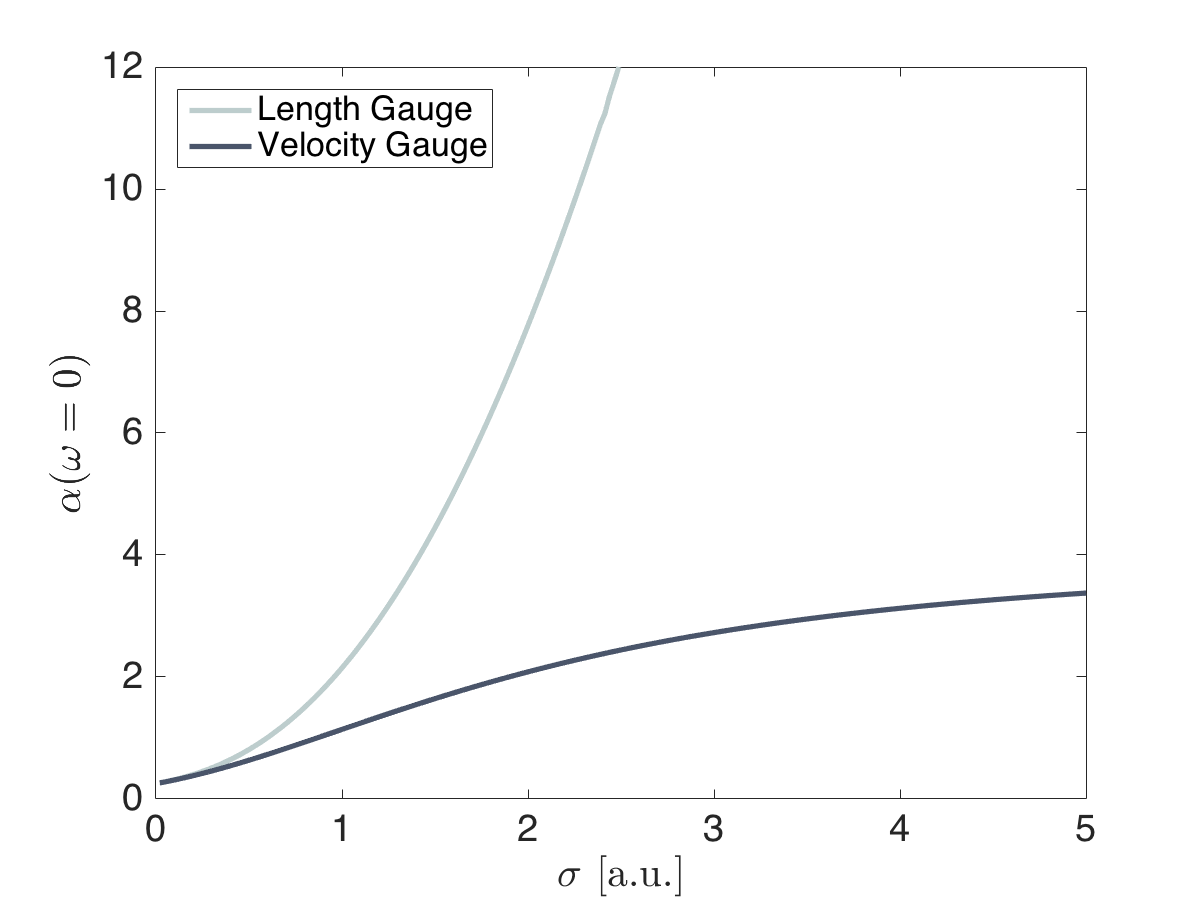}}
\caption{(Color online) The static polarizability $\alpha(\omega = 0)$ as calculated from Eqs. \eqref{eq:VG_pol} and \eqref{eq:LG_pol}.  In the limit $\sigma \rightarrow 0$, the nonlocal potential is equivalent to a gaussian potential, and the polarizability is gauge-independent.  For positive values of $\sigma$, the length-gauge system is more easily polarized by a (DC) applied electric field.}
\label{fig:alpha_vs_sigma}
\end{figure}

Equations \eqref{eq:VG_pol} and \eqref{eq:LG_pol} are evaluated in the limit $\omega \rightarrow 0$ in Fig \ref{fig:alpha_vs_sigma} to show the static (DC) polarizability as a function of the fitting parameter $\sigma$.  In the limit $\sigma \rightarrow 0$, the gaussian nonlocal potential is equivalent to an attractive delta function potential, and the polarizability is observed to limit to a non-zero gauge-independent value.  If $\sigma$ is increased, the length gauge static polarizability is observed to be much greater than that in the velocity gauge formulation.  For comparison, the established (non-relativistic) values of the static polarizability for several atomic species are as follows: hydrogen: 4.5[a.u], helium: 1.38[a.u], neon: 2.68[a.u], and argon: 11.10[a.u] \citep{Schwerdtfeger2013}.

\begin{figure}
    \centering
    \begin{subfigure}[h]{.38\textwidth}
        \includegraphics[width=\textwidth]{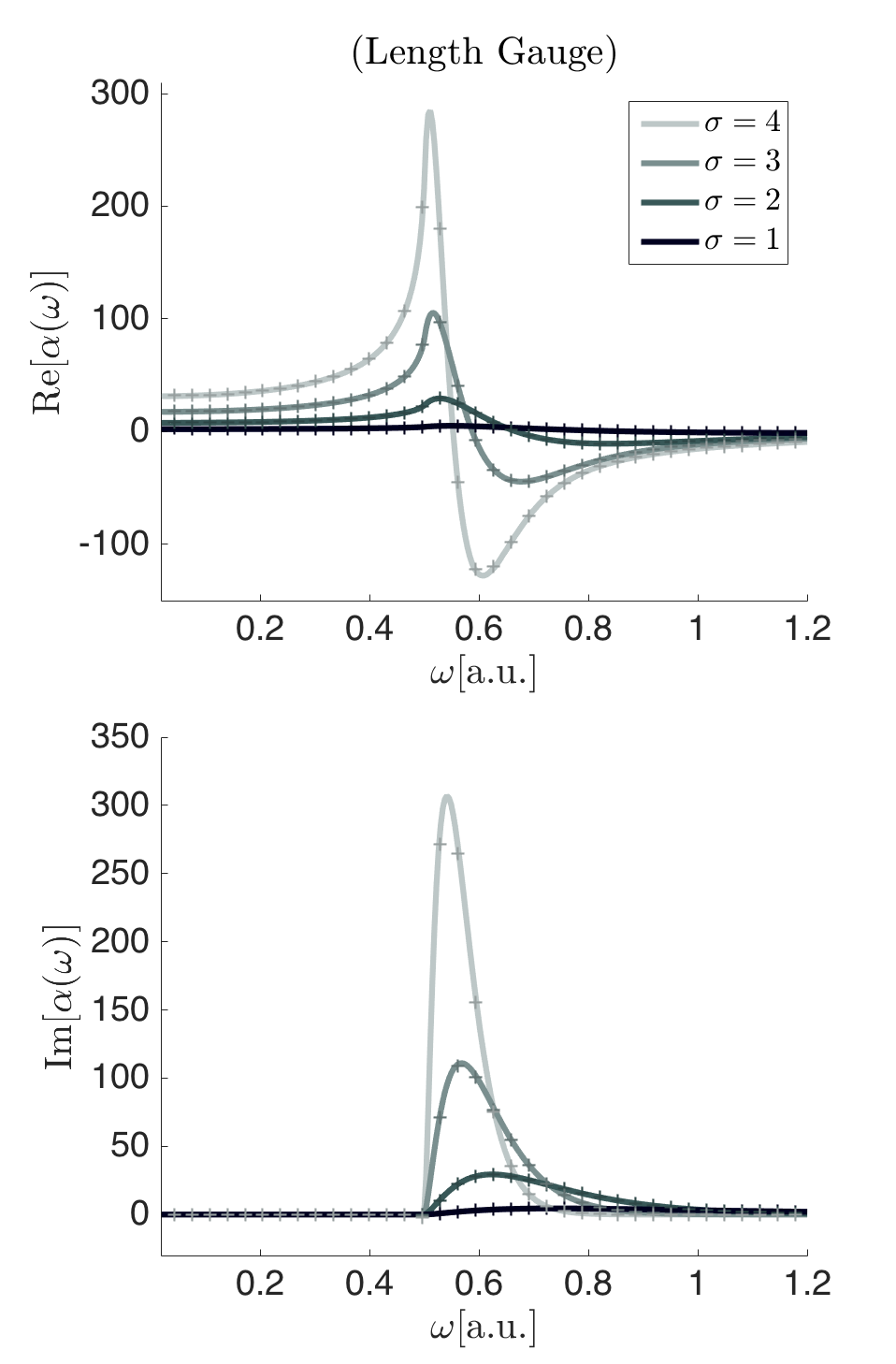}\\[-.02 ex]
        \caption{}
        \label{fig:alpha_LG}
    \end{subfigure}
    \begin{subfigure}[h]{.38\textwidth}
        \includegraphics[width=\textwidth]{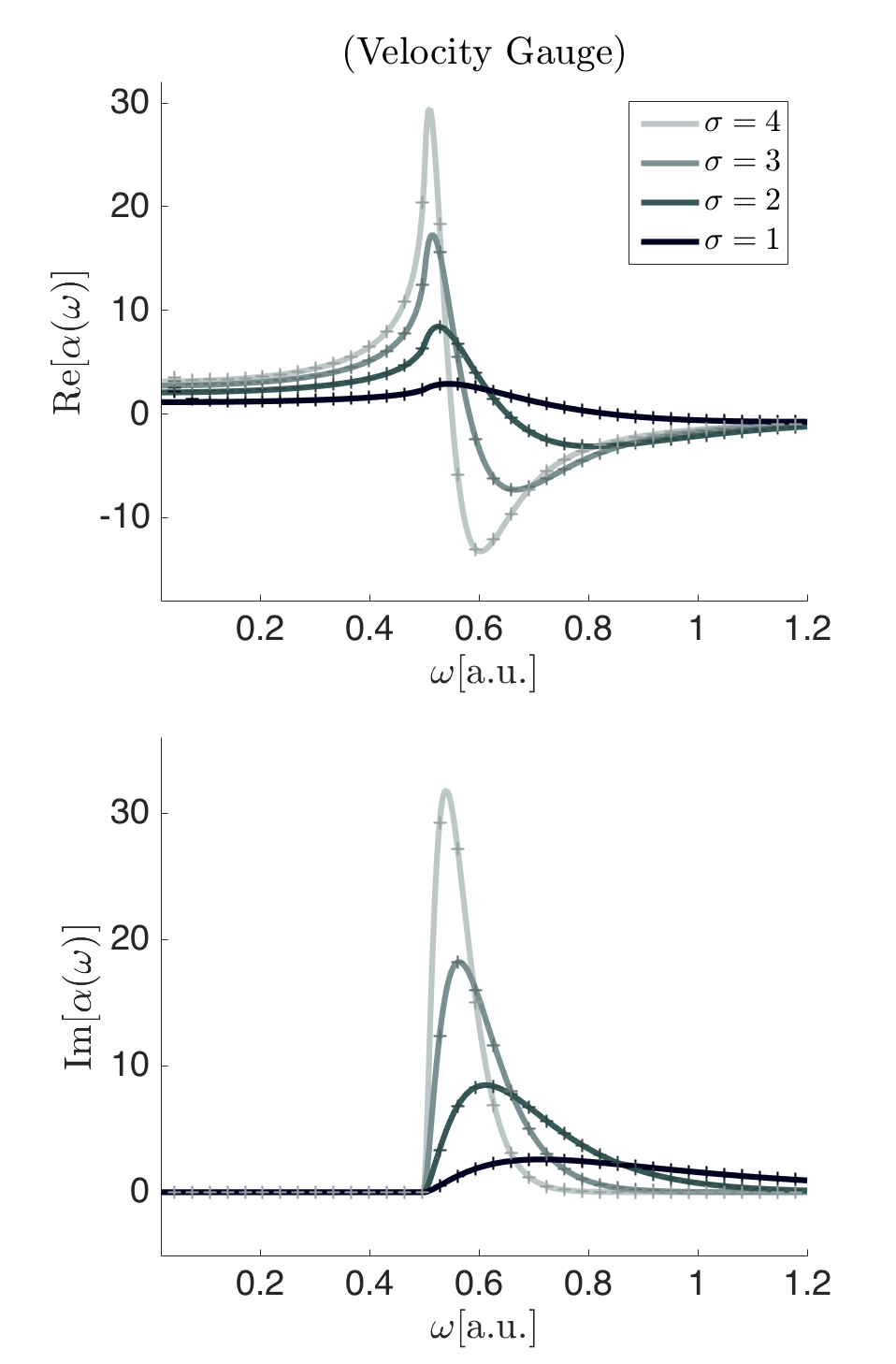}\\[-.02 ex]
        \caption{}
        \label{fig:alpha_VG}
    \end{subfigure}
    \caption{The dynamic polarizability in the length and velocity gauges, with a single photon resonance at $\omega = E_0$  The solid lines represent the $\alpha(\omega)$ given by Eqs.\eqref{eq:VG_pol} and \eqref{eq:LG_pol} ($\omega \geq E_0 \rightarrow \omega + i\delta$), and the crosses represent the simulated low field response via the total dipole (Eq.\eqref{eq:dipole}).}
    \label{fig:alpha_w}
\end{figure}

In Fig.(\ref{fig:alpha_w}), $\alpha(\omega)$ is evaluated via Eqs.\eqref{eq:VG_pol} and \eqref{eq:LG_pol} (solid lines) and plotted as a function of laser frequency for various values of sigma.  The polarization is real for $\omega < E_0$ but complex for $\omega > E_0$.  To evaluate $\alpha(\omega)$ for $\omega \geq E_0$ the laser frequency, previously defined as real, is allowed to become slightly complex, $\omega \rightarrow \omega + i \delta$ accounting for causality.  It should be noted that systems with additional eigenstates would have additional resonances for coupling to excited states, e.g. for hydrogen: $\omega_{res} = E_0(1- 1/n^2)$.  These are not present in a single bound state system.
The cross marks in Fig.\ref{fig:alpha_w} represent the $\alpha(\omega)$ calculated from numerical simulation via Eqs.\eqref{eq:k} and \eqref{eq:dipole}.  Each cross represents the atomic dipole calculated for a 50 femtosecond low intensity pulse ($I_{\text{max}} = 1\times10^{10}$ W/cm$^2$), and the ratio taken of Fourier transform coefficients $\mathbf {\hat p}(\omega), \mathbf {\hat F}(\omega)$.  Agreement is observed between the predicted and simulated values.

\subsection{Ionization}
The time dependent bound-electron probability is defined as
\begin{equation}\label{eq:rho}
\rho(t) \equiv \left| \int d^3 \mathbf x'\ \psi_0^*(\mathbf x') \psi(\mathbf x',t) \right|^2 = \left|\int d^3 \mathbf k' \  \phi_0^*(\mathbf k') \phi(\mathbf k',t)\right|^2;
\end{equation}
we may use this to define the time dependent ionization rate $\nu(t)$ through the relation
\begin{equation}\label{eq:ion_rate}
\rho(t) = \rho(t_0)\exp\left[- \int_{t_0}^t \nu(t')dt' \right],
\end{equation}
that depends functionally on the field $\mathbf F(t)$.
\begin{figure}
\centering
\begin{subfigure}[h]{.48\textwidth}
\includegraphics[width=\textwidth]{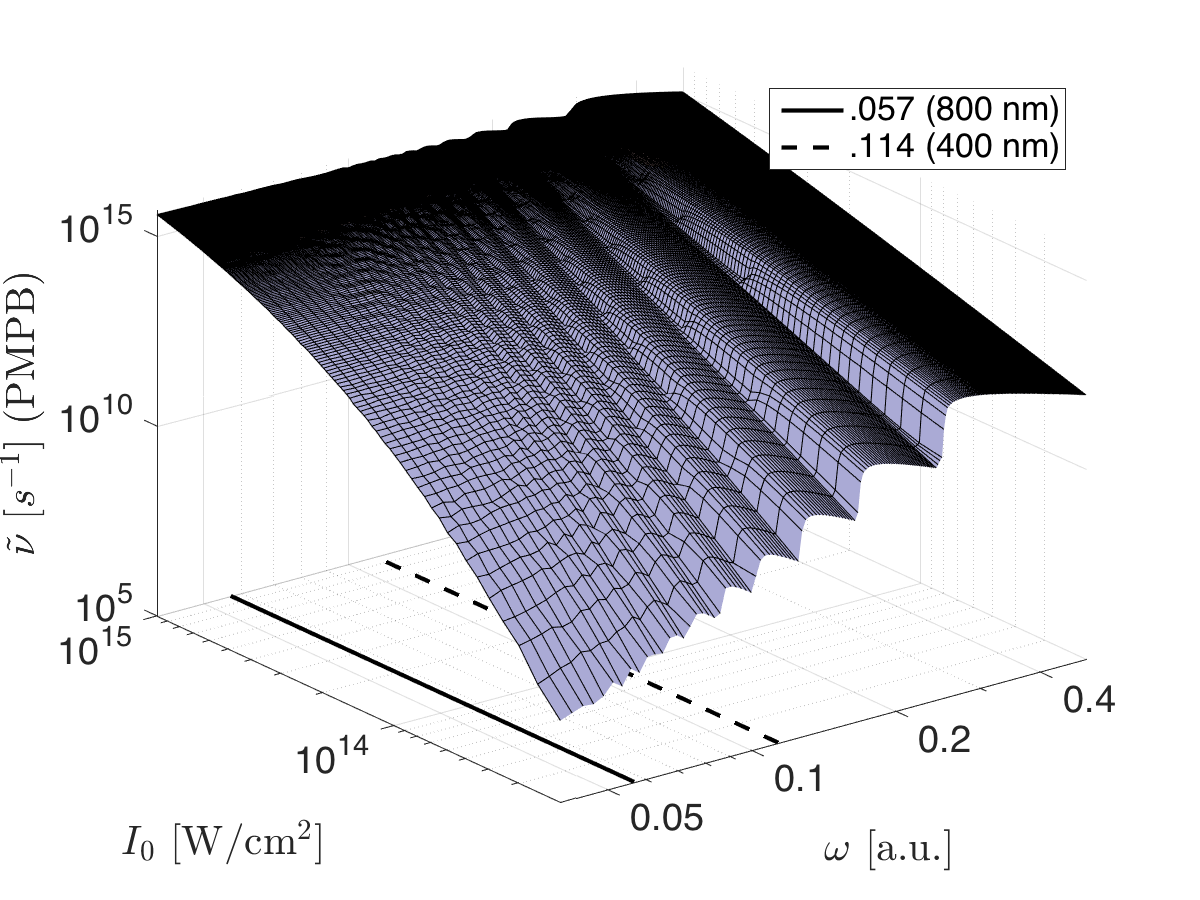}\\[-.02 ex]
\label{fig:nu_PMPB}
\end{subfigure}
\begin{subfigure}[h]{0.48\textwidth}
\includegraphics[width=\textwidth]{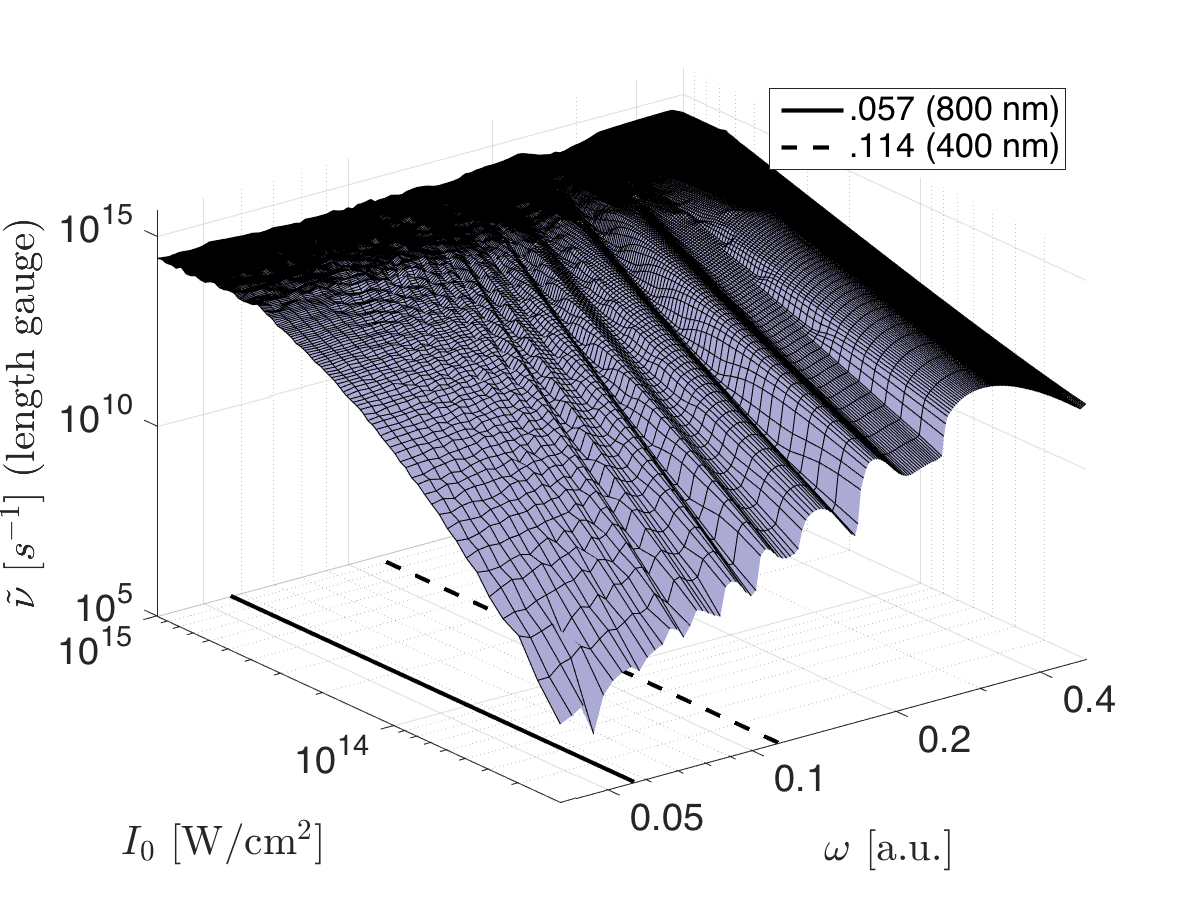}\\[-.02 ex]
\label{fig:nu_LG}
\end{subfigure}
\begin{subfigure}[h]{0.48\textwidth}
\includegraphics[width=\textwidth]{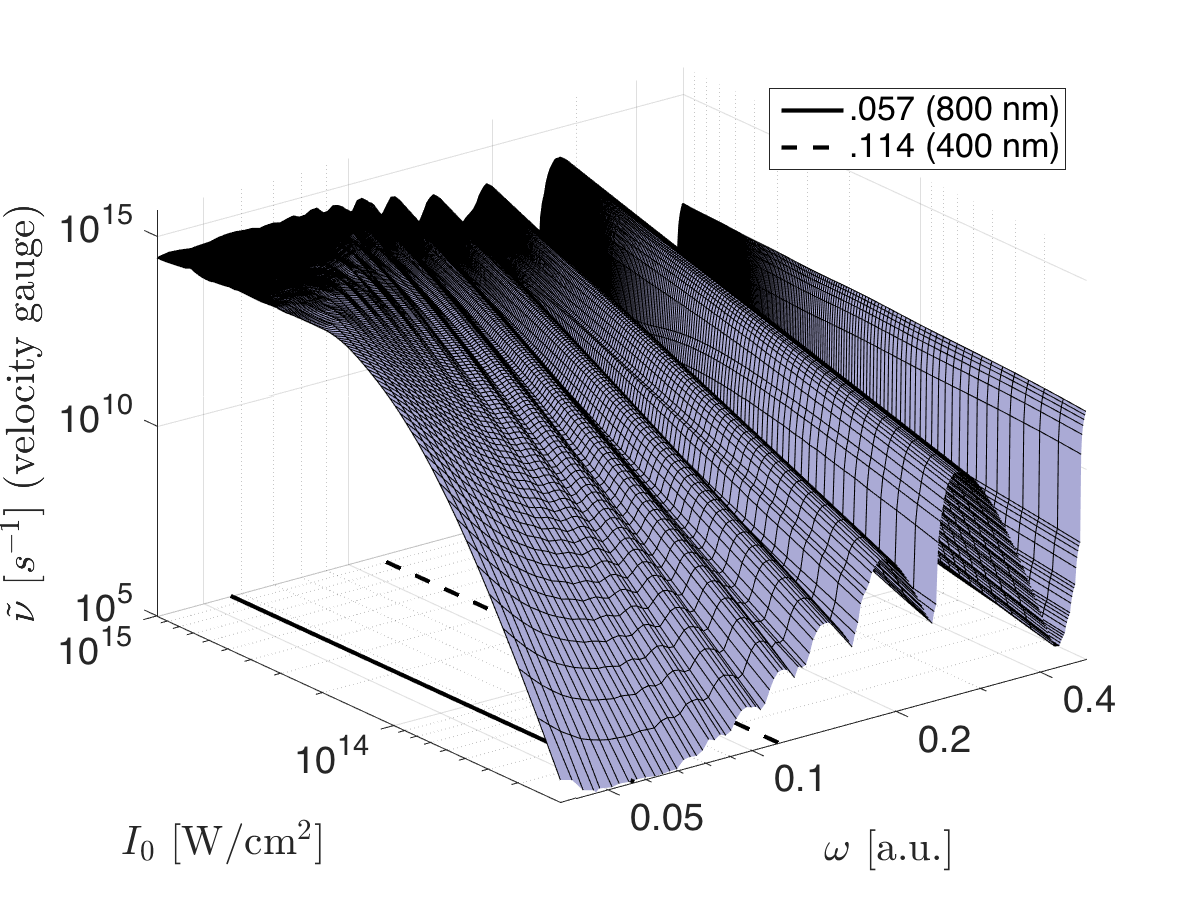}\\[-.02 ex]
\label{fig:nu_VG}
\end{subfigure}
\caption{The ionization rates predicted by the PMPB model and the nonlocal length/velocity gauge formulations as a function of laser frequency and intensity ($100\times100$ data points, interpolated).  The laser parameters here span the multiphoton (high frequency, low intensity) and tunnel (low frequency, high intensity) ionization regimes.  The length gauge and PMPB ionization rates agree well over the parameter space shown; the velocity gauge ionization rate generally underestimates in the multiphoton regime and overestimates in the tunnel regime.  Slices along constant intensity and frequency are shown in Figs.\ref{fig:intensity_dependence} and \ref{fig:frequency_dependence} for direct comparison.}
\label{fig:nu_2D}
\end{figure}

\begin{figure}\label{fig:freq_dep}
\centering
\begin{subfigure}[h]{.45\textwidth}
\includegraphics[width=\textwidth]{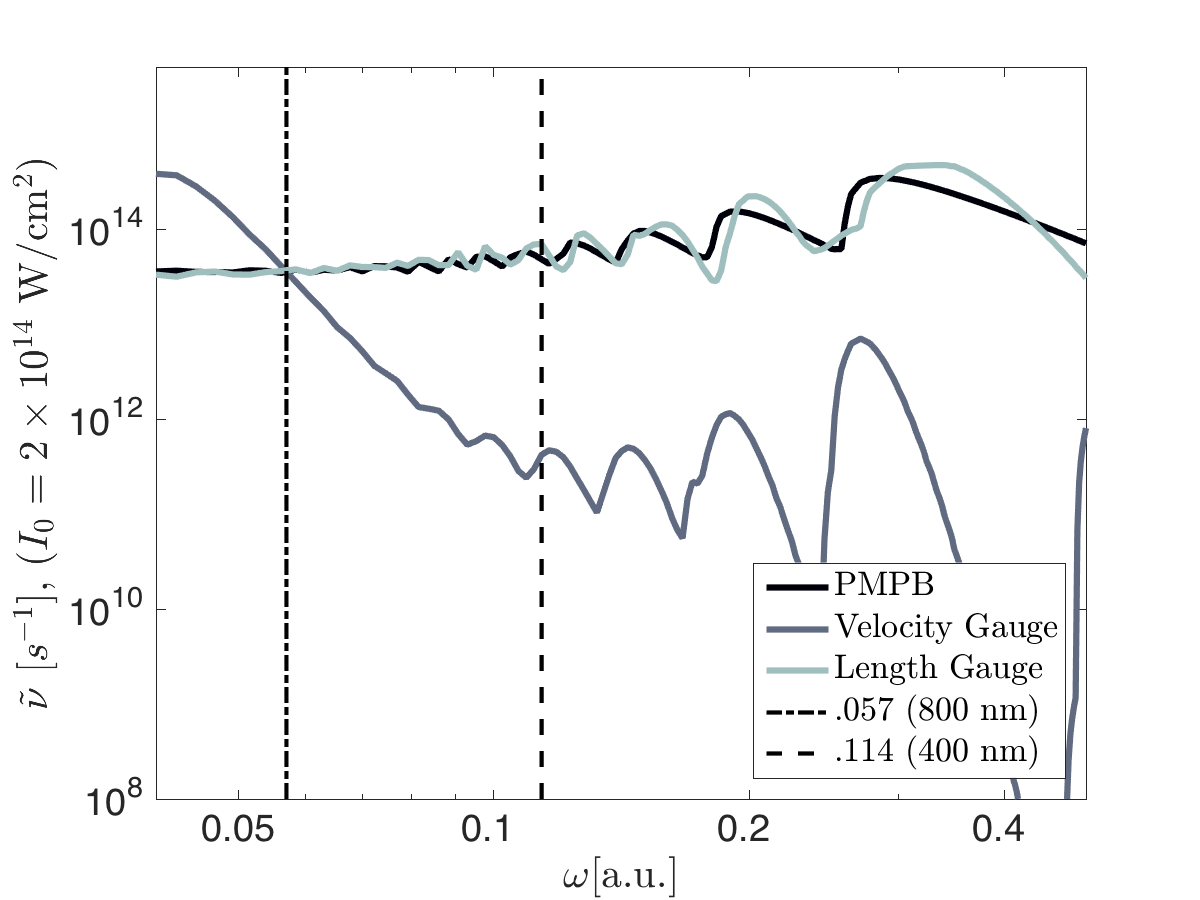}\\[-.02 ex]
\caption{}\label{fig:nu_Wc1_3}
\end{subfigure}
\begin{subfigure}[h]{.45\textwidth}
\includegraphics[width=\textwidth]{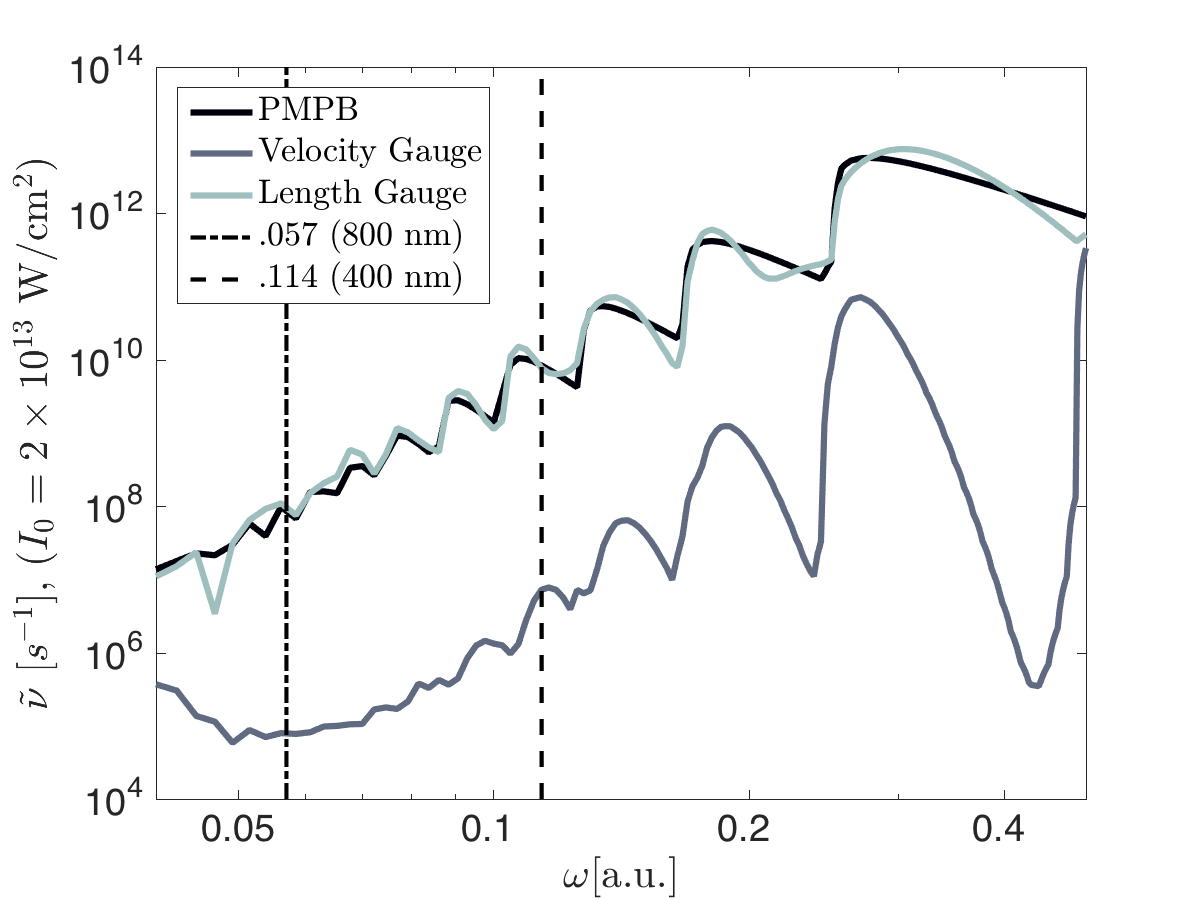}\\[-.02 ex]
\caption{}
\label{fig:nu_Wc1_1}
\end{subfigure}
\caption{The ionization rate as a function of intensity for $I_0=2\times 10^{14}$W/cm$^2$ (\ref{fig:nu_Wc1_3}); the values of $\sigma$ for the length and velocity gauge potentials were calibrated at this intensity, at 800 nm (visible here as the crossing point for all rates).  The velocity gauge overestimates the rate towards tunnel regime and underestimates it in the multiphoton regime, while the PMPB and length gauge rates predict similar rates.  The rates are also plotted for $I_0=2\times 10^{13}$W/cm$^2$ (\ref{fig:nu_Wc1_1}).}
\label{fig:frequency_dependence}
\end{figure}

\begin{figure}
\centering
\begin{subfigure}[h]{.45\textwidth}
\includegraphics[width=\textwidth]{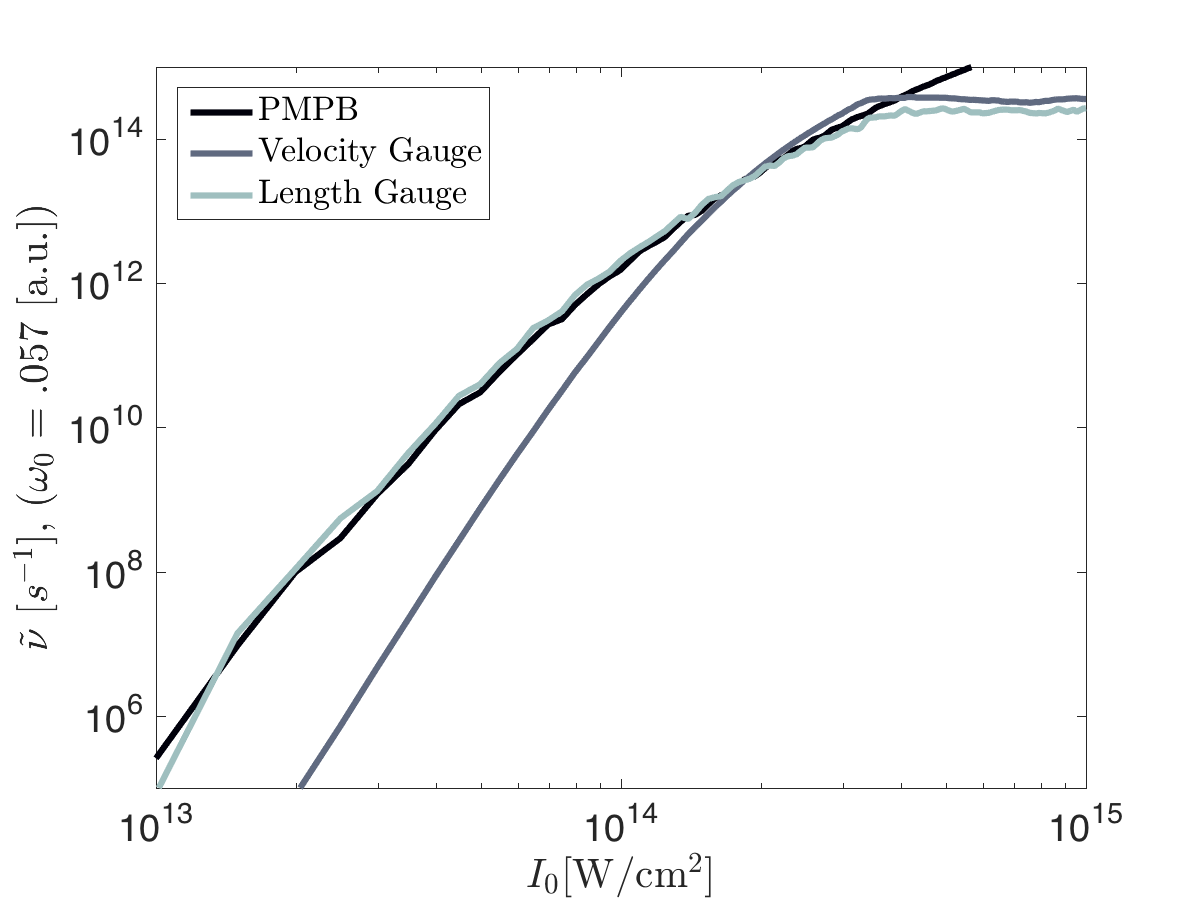}\\[-.02 ex]
\caption{}
\label{fig:nu_I0_1}
\end{subfigure}
\begin{subfigure}[h]{.45\textwidth}
\includegraphics[width=\textwidth]{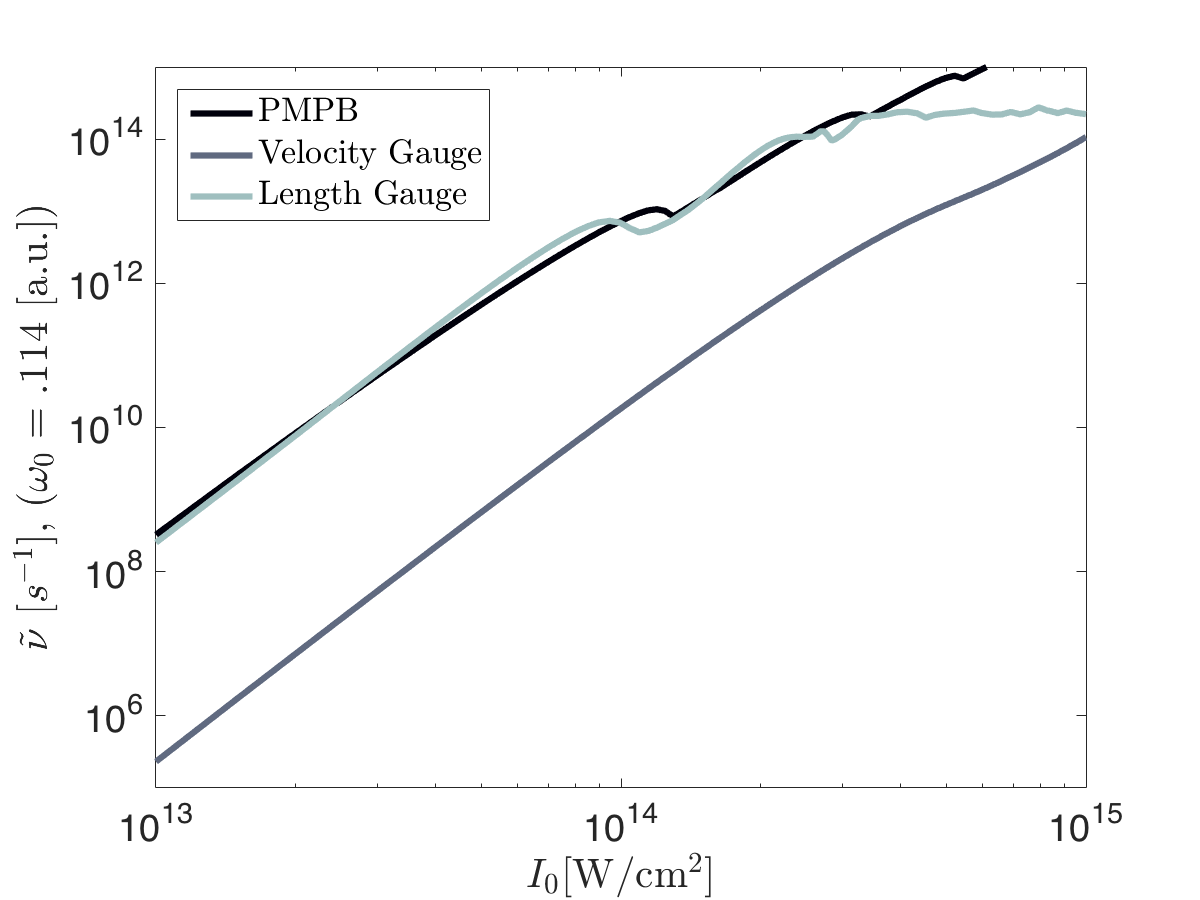}\\[-.02 ex]
\caption{}
\label{fig:nu_I0_2}
\end{subfigure}
\caption{The PMPB photoionization rate and nonlocal (length gauge) rate also show agreement as a function of intensity for 800 nanometer light; the velocity gauge ionization rate does not (\ref{fig:nu_I0_1}).  At 400 nm (\ref{fig:nu_I0_2}), the velocity gauge ionization rate has the same power dependence as the length gauge and PMPB rates, but strongly underestimates the magnitude for the chosen fitting parameter ($\sigma = 4.785$).}
\label{fig:intensity_dependence}
\end{figure}

In practice, it is often easier to use functions other than $\rho(t)$ that are approximately equal to the bound probability defined by Eq. \eqref{eq:rho}.  For the nonlocal potential, we use the quantity 
\begin{equation}\label{eq:rho_tilde}
\rho_u(t) \equiv \left|\frac{S(t)}{S_0}\right|^2 \propto \left|\int d^3 \mathbf k' \  u^*(\mathbf k') \phi(\mathbf k',t)\right|^2
\end{equation}
for convenience, noting that the functions $u(\mathbf k)$ and $\phi_0(\mathbf k)$ are similar in functional form, and note the limit $E_0\rightarrow \infty$, $\rho_u(t)\rightarrow \rho(t)$ .  Although these measures are not identical, any wave density that escapes the nonlocal potential region quickly propagate away from the origin, making $\rho_u(t)$ a very good approximation of the bound probability.  A comparison of these quantities was rather carefully examined in previous work \citep{Rensink2014} which demonstrated $\rho_u(t)$ and $\rho(t)$ were in agreement in the length gauge formulation.  The quantity $\rho_u(t)$ does not offer such a straightforward interpretation in the velocity gauge, but can be used as a measure of bound probability for times when $\mathbf A(t) = 0$, and can be used for measuring pulse averaged ionization rates.

We compare the length and velocity gauge predicted ionization rates using a flat-top laser pulse of form $\mathbf E(t) \equiv F(t)\cos(\omega t)\mathbf{\hat z}$ with a 15 femtosecond ramp-time ($t_r$) of the form
\begin{equation}\label{eq:pulse_def}
F(t) \equiv \left\{
	\begin{array}{ll}
		F_0\sin^2(\frac{\pi}{2t_r}t)   & \mbox{for } 0 \leq t \leq t_r \\
		F_0 & \mbox{for } t_r < t \leq t_p-t_r\\
		F_0\cos^2(\frac{\pi}{2t_r}(t-t_p+t_r))   & \mbox{for } t_p-t_r < t \leq t_p
	\end{array}
\right.
\end{equation}
where $t_r$ is the ramp time to maximum and $t_p$ is the total pulse length, with values of 15 and 90 femtoseconds respectively.  This pulse profile was used in place of a gaussian or $\sin^2(t)$ envelope to maximize the time the electric field amplitude was at a fixed value while still maintaining a narrow bandwidth to prevent frequency dependent structure in the ionization rate from being averaged out.  The total drop in bound probability $\rho_u(t_{f})$ (Eq. \eqref{eq:rho_tilde}) is used to calculate a pulse averaged ionization rate
\begin{equation}\label{eq:nu_ave}
 \tilde \nu = - \frac{ln\left[\rho(t_f) \right]}{t_p - t_r}.
\end{equation}

Figure \ref{fig:nu_2D} shows the ionization rate $\tilde \nu$ landscape as a function of the near infrared to near ultraviolet laser frequency at ionizing intensities, spanning the multiphoton and tunnel ionization regimes.   The length gauge and velocity gauge rates are compared with an ionization rate model introduced by Popruzhenko, et. al. in 2008 \citep{Popruzhenko2008}, here referred to as the ``PMPB'' rate, in reference to the authors' names.  The PMPB rate used for comparison here is preferable to Keldysh or ADK models \citep{Keldysh1964, Ammosov1986, Karnakov2015} which are known to underestimate the multiphoton ionization rate by several orders of magnitude; the PMPB model is valid in both the tunneling and multiphoton regimes and was shown to give good agreement with both Floquet and \textit{ab initio} TDSE solver simulations \citep{Popruzhenko2008}.  

To compare the ionization rate predicted by the nonlocal potential, the tuning parameter $\sigma$ was fixed by matching the ionization rate of a single run with typical laboratory parameters $\omega = .057$ [a.u.] (800 nm), and $F_0 = .01$ [a.u.] (Intensity of $2 \times 10^{14}$ W/cm$^2$), seen as the crossing point of all rates in Fig.\ref{fig:nu_Wc1_3}.  The values $\sigma = 2.482$ for the length gauge and $\sigma = 4.785$ for the velocity gauge were used in all ionization plots shown.

A glance at Fig. \ref{fig:nu_2D} shows that the nonlocal length gauge and PMPB rates share the same general contours across the entire range of intensities and frequencies shown here.  Slices taken along lines of constant frequency (\ref{fig:nu_I0_1}, \ref{fig:nu_I0_2}) and constant intensity (\ref{fig:nu_Wc1_1}, \ref{fig:nu_Wc1_3}) give a more direct comparison and show strong agreement in the PMPB and length gauge ionization rates for all frequencies examined and intensities up to $I_0\sim4\times10^{14}$ W/cm$^2$.  The deviation above this intensity is only apparent; calculation of the $S(t)$ always leaves residual traces which artificially decrease $\tilde \nu$.  The PMPB and nonlocal length gauge predict similar ionization rates for all laser parameters shown.  It should be stated that the agreement in ionization rate shown in these plots is, in some cases misleading; neither the PMPB nor the nonlocal model here can account for ionization pathways that include intermediate population of excited electron states, \citep{Serebryannikov2016}.

By contrast, the velocity gauge ionization rate does not agree with the PMPB rate; it underestimates ionization below $I_0=2\times10^{14}$ W/cm$^2$ and overestimates it for higher intensities; for this reason, it is unlikely that a different choice of $\sigma$ could improve the predicted rate in the tunnel and multiphoton regimes (the rate generally changes monotonically with the tuning parameter $\sigma$ for a specified electric field).  This under-prediction at low intensities and over prediction at high-intensities for the velocity gauge formulation is consistent with other work \citep{TetchouNganso2013} which examined the ionization rate of a similar nonlocal model in the velocity gauge.

\section{CONCLUSION}
In this work, we examine the gauge dependence of nonlocal atomic potentials in the time dependent Schrodinger equation.  We note that the utility of nonlocal potential models is that the atom-field interaction can be computed in the time domain without having to resolve the spatial or momentum space wavefunction, allowing for rapid evaluation of e.g., the atomic dipole moment and photoionization rate.  For this reason, nonlocal potential models are of interest for examining atom-field interactions and for use in Maxwell-Schrodinger laser propagation simulations. 

Specifically, we consider the linear dipolar field response and photoionization rate, predicted by a gaussian nonlocal atomic potential, in the length and velocity gauges in a time varying electric field.  All examined quantities are found to be gauge dependent.  At low intensities (I $\sim10^{10} $W/cm$^2$ and below), both gauge formulations exhibit similar resonant frequency response at photon energies near the ionization threshold, and a static polarizability in the low frequency limit, but differ significantly in magnitude.  The photoionization rates predicted in each gauge were compared with the Coulombic photoionization rate model (PMPB) \citep{Popruzhenko2008}, in the frequency (near IR to near UV) and intensity domains ((I $\sim10^{13} - 10^{15}$W/cm$^2$).  It was found that, although gauge formulations demonstrate multiphoton resonance and tunnel features, the velocity gauge formulation generally over estimated the tunnel ionization rate and underestimated the multiphoton ionization rate; the length gauge and PMPB photoionization rates agreed well over the entire parameter range investigated.

\section*{ACKNOWLEDGEMENT}
This work was supported by the U.S. Department of Energy (DoE) and Naval Research Laboratory (NRL).

\bibliographystyle{apsrev.bst}
\bibliography{/Users/trensink/Dropbox/publication/NLI_gauge_submission01/bibliography/nonlocal_gauge}

\begin{thebibliography}{23}
\expandafter\ifx\csname natexlab\endcsname\relax\def\natexlab#1{#1}\fi
\expandafter\ifx\csname bibnamefont\endcsname\relax
  \def\bibnamefont#1{#1}\fi
\expandafter\ifx\csname bibfnamefont\endcsname\relax
  \def\bibfnamefont#1{#1}\fi
\expandafter\ifx\csname citenamefont\endcsname\relax
  \def\citenamefont#1{#1}\fi
\expandafter\ifx\csname url\endcsname\relax
  \def\url#1{\texttt{#1}}\fi
\expandafter\ifx\csname urlprefix\endcsname\relax\def\urlprefix{URL }\fi
\providecommand{\bibinfo}[2]{#2}
\providecommand{\eprint}[2][]{\url{#2}}

\bibitem[{\citenamefont{Kim et~al.}(2007)\citenamefont{Kim, Glownia, Taylor,
  and Rodriguez}}]{Kim2007}
\bibinfo{author}{\bibfnamefont{K.-Y.} \bibnamefont{Kim}},
  \bibinfo{author}{\bibfnamefont{J.~H.} \bibnamefont{Glownia}},
  \bibinfo{author}{\bibfnamefont{A.~J.} \bibnamefont{Taylor}},
  \bibnamefont{and}
  \bibinfo{author}{\bibfnamefont{G.}~\bibnamefont{Rodriguez}},
  \bibinfo{journal}{Opt. Express} \textbf{\bibinfo{volume}{15}},
  \bibinfo{pages}{4577} (\bibinfo{year}{2007}), ISSN \bibinfo{issn}{1094-4087}.

\bibitem[{\citenamefont{Johnson et~al.}(2013)\citenamefont{Johnson, Palastro,
  Antonsen, and Kim}}]{Johnson2013}
\bibinfo{author}{\bibfnamefont{L.~A.} \bibnamefont{Johnson}},
  \bibinfo{author}{\bibfnamefont{J.~P.} \bibnamefont{Palastro}},
  \bibinfo{author}{\bibfnamefont{T.~M.} \bibnamefont{Antonsen}},
  \bibnamefont{and} \bibinfo{author}{\bibfnamefont{K.~Y.} \bibnamefont{Kim}},
  \bibinfo{journal}{Phys. Rev. A} \textbf{\bibinfo{volume}{88}},
  \bibinfo{pages}{063804} (\bibinfo{year}{2013}), ISSN
  \bibinfo{issn}{1050-2947},
  \urlprefix\url{http://link.aps.org/doi/10.1103/PhysRevA.88.063804}.

\bibitem[{\citenamefont{Chen et~al.}(2015)\citenamefont{Chen, Huang, Meng, Liu,
  Zhou, Zhang, Yuan, and Zhao}}]{Chen2015}
\bibinfo{author}{\bibfnamefont{W.}~\bibnamefont{Chen}},
  \bibinfo{author}{\bibfnamefont{Y.}~\bibnamefont{Huang}},
  \bibinfo{author}{\bibfnamefont{C.}~\bibnamefont{Meng}},
  \bibinfo{author}{\bibfnamefont{J.}~\bibnamefont{Liu}},
  \bibinfo{author}{\bibfnamefont{Z.}~\bibnamefont{Zhou}},
  \bibinfo{author}{\bibfnamefont{D.}~\bibnamefont{Zhang}},
  \bibinfo{author}{\bibfnamefont{J.}~\bibnamefont{Yuan}}, \bibnamefont{and}
  \bibinfo{author}{\bibfnamefont{Z.}~\bibnamefont{Zhao}},
  \bibinfo{journal}{Phys. Rev. A - At. Mol. Opt. Phys.}
  \textbf{\bibinfo{volume}{92}} (\bibinfo{year}{2015}), ISSN
  \bibinfo{issn}{10941622}, \eprint{1503.07588}.

\bibitem[{\citenamefont{Lewenstein et~al.}(1994)\citenamefont{Lewenstein,
  Balcou, Ivanov, L'Huillier, and Corkum}}]{Lewenstein1994}
\bibinfo{author}{\bibfnamefont{M.}~\bibnamefont{Lewenstein}},
  \bibinfo{author}{\bibfnamefont{P.}~\bibnamefont{Balcou}},
  \bibinfo{author}{\bibfnamefont{M.~Y.} \bibnamefont{Ivanov}},
  \bibinfo{author}{\bibfnamefont{A.}~\bibnamefont{L'Huillier}},
  \bibnamefont{and} \bibinfo{author}{\bibfnamefont{P.~B.}
  \bibnamefont{Corkum}}, \bibinfo{journal}{Phys. Rev. A}
  \textbf{\bibinfo{volume}{49}}, \bibinfo{pages}{2117} (\bibinfo{year}{1994}),
  ISSN \bibinfo{issn}{10502947}, \eprint{1106.1603},
  \urlprefix\url{http://link.aps.org/doi/10.1103/PhysRevA.49.2117}.

\bibitem[{\citenamefont{Dahlstrom et~al.}(2013)\citenamefont{Dahlstrom, Guenot,
  Klunder, Gisselbrecht, Mauritsson, L'Huillier, Maquet, and
  Taieb}}]{Dahlstrom2013}
\bibinfo{author}{\bibfnamefont{J.~M.} \bibnamefont{Dahlstrom}},
  \bibinfo{author}{\bibfnamefont{D.}~\bibnamefont{Guenot}},
  \bibinfo{author}{\bibfnamefont{K.}~\bibnamefont{Klunder}},
  \bibinfo{author}{\bibfnamefont{M.}~\bibnamefont{Gisselbrecht}},
  \bibinfo{author}{\bibfnamefont{J.}~\bibnamefont{Mauritsson}},
  \bibinfo{author}{\bibfnamefont{A.}~\bibnamefont{L'Huillier}},
  \bibinfo{author}{\bibfnamefont{A.}~\bibnamefont{Maquet}}, \bibnamefont{and}
  \bibinfo{author}{\bibfnamefont{R.}~\bibnamefont{Taieb}},
  \bibinfo{journal}{Chem. Phys.} \textbf{\bibinfo{volume}{414}},
  \bibinfo{pages}{53} (\bibinfo{year}{2013}), ISSN \bibinfo{issn}{03010104},
  \eprint{1112.4144}.

\bibitem[{\citenamefont{Couairon et~al.}(2011)\citenamefont{Couairon,
  Brambilla, Corti, Majus, Ram{\'{i}}rez-G{\'{o}}ngora, and
  Kolesik}}]{Couairon2011}
\bibinfo{author}{\bibfnamefont{A.}~\bibnamefont{Couairon}},
  \bibinfo{author}{\bibfnamefont{E.}~\bibnamefont{Brambilla}},
  \bibinfo{author}{\bibfnamefont{T.}~\bibnamefont{Corti}},
  \bibinfo{author}{\bibfnamefont{D.}~\bibnamefont{Majus}},
  \bibinfo{author}{\bibfnamefont{O.}~\bibnamefont{Ram{\'{i}}rez-G{\'{o}}ngora}},
  \bibnamefont{and} \bibinfo{author}{\bibfnamefont{M.}~\bibnamefont{Kolesik}},
  \bibinfo{journal}{Eur. Phys. J. Spec. Top.} \textbf{\bibinfo{volume}{199}},
  \bibinfo{pages}{5} (\bibinfo{year}{2011}), ISSN \bibinfo{issn}{1951-6355},
  \urlprefix\url{http://www.springerlink.com/index/10.1140/epjst/e2011-01503-3}.

\bibitem[{\citenamefont{Kolesik et~al.}(2016)\citenamefont{Kolesik, Brown, and
  Bahl}}]{Kolesik2016}
\bibinfo{author}{\bibfnamefont{M.}~\bibnamefont{Kolesik}},
  \bibinfo{author}{\bibfnamefont{J.}~\bibnamefont{Brown}}, \bibnamefont{and}
  \bibinfo{author}{\bibfnamefont{A.}~\bibnamefont{Bahl}}
  (\bibinfo{publisher}{International Society for Optics and Photonics},
  \bibinfo{year}{2016}), p. \bibinfo{pages}{983510},
  \urlprefix\url{http://proceedings.spiedigitallibrary.org/proceeding.aspx?doi=10.1117/12.2223428}.

\bibitem[{\citenamefont{Popruzhenko et~al.}(2008)\citenamefont{Popruzhenko,
  Mur, Popov, and Bauer}}]{Popruzhenko2008}
\bibinfo{author}{\bibfnamefont{S.}~\bibnamefont{Popruzhenko}},
  \bibinfo{author}{\bibfnamefont{V.}~\bibnamefont{Mur}},
  \bibinfo{author}{\bibfnamefont{V.}~\bibnamefont{Popov}}, \bibnamefont{and}
  \bibinfo{author}{\bibfnamefont{D.}~\bibnamefont{Bauer}},
  \bibinfo{journal}{Phys. Rev. Lett.} \textbf{\bibinfo{volume}{101}},
  \bibinfo{pages}{193003} (\bibinfo{year}{2008}), ISSN
  \bibinfo{issn}{0031-9007}.

\bibitem[{\citenamefont{Gordon and Hafizi}(2012)}]{Gordon2012}
\bibinfo{author}{\bibfnamefont{D.}~\bibnamefont{Gordon}} \bibnamefont{and}
  \bibinfo{author}{\bibfnamefont{B.}~\bibnamefont{Hafizi}},
  \bibinfo{journal}{J. Comput. Phys.} \textbf{\bibinfo{volume}{231}},
  \bibinfo{pages}{6349} (\bibinfo{year}{2012}), ISSN \bibinfo{issn}{00219991}.

\bibitem[{\citenamefont{Potvliege}(1998)}]{Potvliege1998}
\bibinfo{author}{\bibfnamefont{R.}~\bibnamefont{Potvliege}},
  \bibinfo{journal}{Comput. Phys. Commun.} \textbf{\bibinfo{volume}{114}},
  \bibinfo{pages}{42} (\bibinfo{year}{1998}), ISSN \bibinfo{issn}{00104655},
  \urlprefix\url{http://www.sciencedirect.com/science/article/pii/S0010465598000733}.

\bibitem[{\citenamefont{Lorin et~al.}(2012)\citenamefont{Lorin, Chelkowski,
  Zaoui, and Bandrauk}}]{Lorin2012}
\bibinfo{author}{\bibfnamefont{E.}~\bibnamefont{Lorin}},
  \bibinfo{author}{\bibfnamefont{S.}~\bibnamefont{Chelkowski}},
  \bibinfo{author}{\bibfnamefont{E.}~\bibnamefont{Zaoui}}, \bibnamefont{and}
  \bibinfo{author}{\bibfnamefont{A.}~\bibnamefont{Bandrauk}},
  \bibinfo{journal}{Phys. D Nonlinear Phenom.} \textbf{\bibinfo{volume}{241}},
  \bibinfo{pages}{1059} (\bibinfo{year}{2012}), ISSN \bibinfo{issn}{01672789},
  \urlprefix\url{http://dx.doi.org/10.1016/j.physd.2012.02.013}.

\bibitem[{\citenamefont{Han and Madsen}(2010)}]{Han2010}
\bibinfo{author}{\bibfnamefont{Y.~C.} \bibnamefont{Han}} \bibnamefont{and}
  \bibinfo{author}{\bibfnamefont{L.~B.} \bibnamefont{Madsen}},
  \bibinfo{journal}{Phys. Rev. A - At. Mol. Opt. Phys.}
  \textbf{\bibinfo{volume}{81}}, \bibinfo{pages}{063430}
  (\bibinfo{year}{2010}), ISSN \bibinfo{issn}{10502947},
  \urlprefix\url{http://link.aps.org/doi/10.1103/PhysRevA.81.063430}.

\bibitem[{\citenamefont{{Tetchou Nganso} et~al.}(2007)\citenamefont{{Tetchou
  Nganso}, Giraud, Piraux, Popov, and {Kwato Njock}}}]{TetchouNganso2007}
\bibinfo{author}{\bibfnamefont{H.~M.} \bibnamefont{{Tetchou Nganso}}},
  \bibinfo{author}{\bibfnamefont{S.}~\bibnamefont{Giraud}},
  \bibinfo{author}{\bibfnamefont{B.}~\bibnamefont{Piraux}},
  \bibinfo{author}{\bibfnamefont{Y.~V.} \bibnamefont{Popov}}, \bibnamefont{and}
  \bibinfo{author}{\bibfnamefont{M.~G.} \bibnamefont{{Kwato Njock}}},
  \bibinfo{journal}{J. Electron Spectros. Relat. Phenomena}
  \textbf{\bibinfo{volume}{161}}, \bibinfo{pages}{178} (\bibinfo{year}{2007}),
  ISSN \bibinfo{issn}{03682048}.

\bibitem[{\citenamefont{{Tetchou Nganso} et~al.}(2011)\citenamefont{{Tetchou
  Nganso}, Popov, Piraux, Madro{\~{n}}ero, and Njock}}]{TetchouNganso2011}
\bibinfo{author}{\bibfnamefont{H.~M.} \bibnamefont{{Tetchou Nganso}}},
  \bibinfo{author}{\bibfnamefont{Y.~V.} \bibnamefont{Popov}},
  \bibinfo{author}{\bibfnamefont{B.}~\bibnamefont{Piraux}},
  \bibinfo{author}{\bibfnamefont{J.}~\bibnamefont{Madro{\~{n}}ero}},
  \bibnamefont{and} \bibinfo{author}{\bibfnamefont{M.~G.~K.}
  \bibnamefont{Njock}}, \bibinfo{journal}{Phys. Rev. A}
  \textbf{\bibinfo{volume}{83}}, \bibinfo{pages}{013401}
  (\bibinfo{year}{2011}), ISSN \bibinfo{issn}{1050-2947}.

\bibitem[{\citenamefont{{Tetchou Nganso} et~al.}(2013)\citenamefont{{Tetchou
  Nganso}, Hamido, {Kwato Njock}, Popov, and Piraux}}]{TetchouNganso2013}
\bibinfo{author}{\bibfnamefont{H.~M.} \bibnamefont{{Tetchou Nganso}}},
  \bibinfo{author}{\bibfnamefont{A.}~\bibnamefont{Hamido}},
  \bibinfo{author}{\bibfnamefont{M.~G.} \bibnamefont{{Kwato Njock}}},
  \bibinfo{author}{\bibfnamefont{Y.~V.} \bibnamefont{Popov}}, \bibnamefont{and}
  \bibinfo{author}{\bibfnamefont{B.}~\bibnamefont{Piraux}},
  \bibinfo{journal}{Phys. Rev. A - At. Mol. Opt. Phys.}
  \textbf{\bibinfo{volume}{87}}, \bibinfo{pages}{1} (\bibinfo{year}{2013}),
  ISSN \bibinfo{issn}{10502947}.

\bibitem[{\citenamefont{Rensink et~al.}(2014)\citenamefont{Rensink, Antonsen,
  Palastro, and Gordon}}]{Rensink2014}
\bibinfo{author}{\bibfnamefont{T.~C.} \bibnamefont{Rensink}},
  \bibinfo{author}{\bibfnamefont{T.~M.} \bibnamefont{Antonsen}},
  \bibinfo{author}{\bibfnamefont{J.~P.} \bibnamefont{Palastro}},
  \bibnamefont{and} \bibinfo{author}{\bibfnamefont{D.~F.}
  \bibnamefont{Gordon}}, \bibinfo{journal}{Phys. Rev. A}
  \textbf{\bibinfo{volume}{89}}, \bibinfo{pages}{033418}
  (\bibinfo{year}{2014}), ISSN \bibinfo{issn}{1050-2947},
  \urlprefix\url{http://link.aps.org/doi/10.1103/PhysRevA.89.033418}.

\bibitem[{\citenamefont{Lai}(1983)}]{Lai83}
\bibinfo{author}{\bibfnamefont{C.~S.} \bibnamefont{Lai}}, \bibinfo{journal}{J.
  Phys. A Math. Gen.} \textbf{\bibinfo{volume}{16}}, \bibinfo{pages}{181}
  (\bibinfo{year}{1983}),
  \urlprefix\url{http://iopscience.iop.org/0305-4470/16/6/002}.

\bibitem[{\citenamefont{Boyd}(2007)}]{Boyd2007}
\bibinfo{author}{\bibfnamefont{R.~W.} \bibnamefont{Boyd}},
  \emph{\bibinfo{title}{{Nonlinear Optics}}} (\bibinfo{year}{2007}).

\bibitem[{\citenamefont{Schwerdtfeger}(2013)}]{Schwerdtfeger2013}
\bibinfo{author}{\bibfnamefont{P.}~\bibnamefont{Schwerdtfeger}}
  (\bibinfo{year}{2013}).

\bibitem[{\citenamefont{Keldysh}(1964)}]{Keldysh1964}
\bibinfo{author}{\bibfnamefont{L.~V.} \bibnamefont{Keldysh}},
  \bibinfo{journal}{Sov. Phys. J. Experimantal Theorectical Phys.}
  \textbf{\bibinfo{volume}{47}}, \bibinfo{pages}{1945} (\bibinfo{year}{1964}),
  ISSN \bibinfo{issn}{00385646}.

\bibitem[{\citenamefont{Ammosov et~al.}(1986)\citenamefont{Ammosov, Delone, and
  Krainov}}]{Ammosov1986}
\bibinfo{author}{\bibfnamefont{M.~V.} \bibnamefont{Ammosov}},
  \bibinfo{author}{\bibfnamefont{N.~B.} \bibnamefont{Delone}},
  \bibnamefont{and} \bibinfo{author}{\bibfnamefont{V.~P.}
  \bibnamefont{Krainov}}, \bibinfo{journal}{JetP}
  \textbf{\bibinfo{volume}{64}}, \bibinfo{pages}{1191} (\bibinfo{year}{1986}).

\bibitem[{\citenamefont{Karnakov et~al.}(2015)\citenamefont{Karnakov, Mur,
  Popruzhenko, and Popov}}]{Karnakov2015}
\bibinfo{author}{\bibfnamefont{B.~M.} \bibnamefont{Karnakov}},
  \bibinfo{author}{\bibfnamefont{V.~D.} \bibnamefont{Mur}},
  \bibinfo{author}{\bibfnamefont{S.~V.} \bibnamefont{Popruzhenko}},
  \bibnamefont{and} \bibinfo{author}{\bibfnamefont{V.~S.} \bibnamefont{Popov}},
  \bibinfo{journal}{Physics-Uspekhi} \textbf{\bibinfo{volume}{58}},
  \bibinfo{pages}{3} (\bibinfo{year}{2015}), ISSN \bibinfo{issn}{1063-7869},
  \urlprefix\url{http://iopscience.iop.org/article/10.3367/UFNe.0185.201501b.0003}.

\bibitem[{\citenamefont{Serebryannikov and
  Zheltikov}(2016)}]{Serebryannikov2016}
\bibinfo{author}{\bibfnamefont{E.~E.} \bibnamefont{Serebryannikov}}
  \bibnamefont{and} \bibinfo{author}{\bibfnamefont{a.~M.}
  \bibnamefont{Zheltikov}}, \bibinfo{journal}{Phys. Rev. Lett.}
  \textbf{\bibinfo{volume}{116}}, \bibinfo{pages}{123901}
  (\bibinfo{year}{2016}), ISSN \bibinfo{issn}{0031-9007},
  \urlprefix\url{http://journals.aps.org/prl/abstract/10.1103/PhysRevLett.116.123901}.

\end{thebibliography}
\end{document}